\pdfoutput=1
\documentclass[
	a4paper,
	rgb,dvipsnames,x11names, 
	]{bmcart}[2014/01/24]
\usepackage{amsfonts}
\usepackage[cmex10]{amsmath}
\usepackage{amssymb}
\usepackage{mathrsfs}
\usepackage[nointegrals]{wasysym}
\usepackage{graphicx}
\usepackage[authoryear]{natbib}
\usepackage[backref]{enotez}
\usepackage{hyperref}
\usepackage{bookmark}
\usepackage{datetime}
\usepackage[utf8]{inputenc}

\providecommand{\now}{\protect\formatdate{07}{05}{2019}}

\DeclareGraphicsExtensions{.mps}
\DeclareTranslation{English}{enotez-title}{Endnotes}
\DeclareInstance{enotez-list}{OPSOUSN}{paragraph}{
	heading   = \section*{#1},
	notes-sep = 0pt,
	format    = \indent\normalfont,
	number    = \textsuperscript{#1}
	}

\startlocaldefs
\providecommand{\glFnc}[1]{{\mathnormal{#1}}}
\providecommand{\glSFnc}[1]{{\mathrm{#1}}}
\newcommand{\fcaIR}{\mathtt{I}}
\newcommand{\plNStR}{\Omega_{r}}
\newcommand{\plNStJ}{\Omega_{j}}
\newcommand{\plNVCR}{\upsilon_{\mathrm{r}}}
\newcommand{\plNVCJ}{\upsilon_{\mathrm{j}}}
\newcommand{\plDrSR}{{\widetilde{\varepsilon}}_{\mathrm{r}}}
\newcommand{\plDrSJ}{{\widetilde{\varepsilon}}_{\mathrm{j}}}
\newcommand{\plZedR}{{\mathfrak{z}}_{\mathrm{r}}}
\newcommand{\plZedJ}{{\mathfrak{z}}_{\mathrm{j}}}
\DeclareMathOperator{\stSu}{Su}
\DeclareMathOperator{\stDr}{\Lambda}
\DeclareMathOperator{\glVa}{Va}
\DeclareMathOperator{\glWg}{\glFnc{w}}
\DeclareMathOperator{\glHm}{\glSFnc{h}}
\DeclareMathOperator{\glEta}{\glFnc{H}}
\DeclareMathOperator{\plDrR}{\widetilde{\Lambda}_{\mathrm{r}}}
\DeclareMathOperator{\plDrJ}{\widetilde{\Lambda}_{\mathrm{j}}}
\DeclareMathOperator{\sfhzeta}{\zeta}
\DeclareMathOperator{\sfWitten}{\mathcal{W}}
\DeclareMathOperator{\sfLerch}{\Phi}
\DeclareMathOperator{\sfWittenLerch}{{\mathcal{W}}_{\Phi}}%

\newcommand{\glprec}{\preccurlyeq}
\newcommand{\glimplies}{\rightarrow}

\endlocaldefs

\renewcommand{\doi}[1]{\href{https://doi.org/\detokenize{#1}}{\texttt{https://doi.org/#1}}}

\begin{document}

\begin{frontmatter}

\begin{fmbox}
\dochead{Research\hfill\now}

\title{On the Perturbation of\protect\\Self-Organized Urban Street Networks}

\author[
	addressref={NYUAD},
	corref={NYUAD},
	email={jerome.benoit@nyu.edu}
	]{\inits{JGMB}\fnm{J{\'e}r{\^o}me GM} \snm{Benoit}} 
\author[
	addressref={NYUAD,NYUTSE},
	email={sej7@nyu.edu}
	]{\inits{SEGJ}\fnm{Saif Eddin G} \snm{Jabari}} 

\address[id=NYUAD]{%
	\orgname{New York University Abu~Dhabi},
	\street{Saadiyat Island},
	\postcode{POB 129188},
	\city{Abu~Dhabi},
	\cny{UAE}
	}
\address[id=NYUTSE]{%
	\orgname{New York University Tandon~School~of~Engineering},
	\street{Brooklyn},
	\postcode{NY 11201},
	\city{New~York},
	\cny{USA}
	}

\end{fmbox}

\begin{abstractbox}

\begin{abstract}
We investigate urban street networks as a whole
within the frameworks of information physics and statistical physics.
Urban street networks are envisaged as evolving social systems subject to
a Boltzmann-mesoscopic entropy conservation.
For self-organized urban street networks,
our paradigm has already allowed us
to recover the
effectively observed
scale-free distribution of roads
and to foresee the distribution of junctions.
The entropy conservation is interpreted as
the conservation of the surprisal of the city-dwellers for their urban street network.
In view to extend our investigations to other urban street networks,
we consider to perturb our model
for self-organized urban street networks
by adding an external surprisal drift.
We obtain the statistics
for slightly drifted self-organized urban street networks.
Besides being practical and manageable,
this statistics separates
the macroscopic evolution scale parameter
from the mesoscopic social parameters.
This opens the door to observational investigations
on the universality of the evolution scale parameter.
Ultimately,
we argue that the strength of the external surprisal drift
might be an indicator for the disengagement of the city-dwellers for their city.
\end{abstract}

\begin{keyword}
\kwd{Urban street networks}
\kwd{Self-organizing networks}
\kwd{Entropic equilibrium}
\kwd{MaxEnt}
\kwd{power law}
\kwd{City science}
\kwd{Interdisciplinary physics}
\kwd{Information physics}
\kwd{Statistical physics}
\kwd{Surprisal}
\kwd{Wholeness}
\kwd{Big data}
\end{keyword}

\end{abstractbox}

\end{frontmatter}

\hypersetup{%
	pdfdisplaydoctitle=true,
	pdftitle={On the Perturbation of Self-Organized Urban Street Networks},
	pdfauthor={J\'er\^ome~Benoit (ORCID: 0000-0003-1226-6757) and Saif~Eddin~Jabari (ORCID: 0000-0002-2314-5312)},
	pdfsubject={%
		Applied Network Science special issue:
		ComplexNetwork 2018
		(%
			The 7th International Conference on Complex Networks and Their Applications
			- Cambridge 11-13 December (United Kingdom)%
			)
		},
	pdfcreator={\LaTeXe{} and its friends},
	pdfpagelayout=SinglePage,
	pdfpagemode=UseOutlines,
	pdfstartpage=1,
	pdfhighlight=/O,
	pdfview=FitH,
	pdfstartview=FitH,
	colorlinks=true,
	allcolors=RedOrange,
	citecolor=RoyalBlue3,
	urlcolor=RoyalBlue3,
	linkcolor=RoyalBlue3,
	bookmarksnumbered=true,
	bookmarksopen=true,
	bookmarksopenlevel=3,
	}

\section*{\addcontentsline{toc}{section}{Introduction}Introduction}

We seek to understand the statistics of urban street networks.
Such an understanding will help urban designers and decision makers
to improve urban policies in general and urban transportation in particular.
In our work we investigate urban street networks as a whole
within the frameworks of information physics \citep{KHKnuth2011}
and statistical physics \citep{ETJaynes1957I,ETJaynes1957II}.

Although
the number of times that a \emph{natural road} crosses an other one
has been widely observed to follow a
discrete Pareto probability distribution \citep{AClausetCRShaliziMEJNewman2009}
among self-organized cities \citep{CAlexanderACINAT1965,CrucittiCMSNUS2006,BJiangTSUSNPDC2014},
very few efforts have focused
on deriving
the statistics of urban street networks
from fundamental principles.
Here a natural road (or road) denotes an accepted substitute for a ``named'' street \citep{BJiangTSUSNPDC2014}.
In a recent work \citep{SESOPLUSN},
we introduce a statistical physics model that derives
the statistics of self-organized urban street networks
by applying Jaynes’s \emph{maximum entropy principle} \citep{ETJaynes1957I,ETJaynes1957II}
through the information physics paradigm \citep{KHKnuth2011}.

Our approach explicitly emphasizes
the road-junction hierarchy of the initial urban street network
rather than implicitly splitting it accordingly in two dual but distinct networks.
Most of the investigations indeed seek
to cast the initial urban street network into a road-road
topological
network \citep{BJiangTSUSNPDC2014}
and to describe its valence probability distribution.
This holistic viewpoint adopted by the urban community \citep{CAlexanderACINAT1965,RHAtkin1974}
appears to fit well with the mindset of information physics \citep{KHKnuth2011},
which is built upon partial order relations \citep{BADaveyHAPriestleyILO,KHKnuth2011}.
Here
the partial order relation derives from the road-junction incidence relation.
The passage from the road-junction hierarchy to a Paretian coherence occurs
by imposing a Boltzmann-mesoscopic entropy conservation \citep{MMilakovic2001,YDover2004}.
The emerging statistical physics expresses better in terms of surprisal \citep{MTribusTT}.
Surprisal quantifies our astonishment and indecision whenever we face an arbitrary event.
Here surprisal betrays the perception of the city-dwellers for their own urban street network.
Then,
the passage to Paretian coherence simply expresses
the conservation on average of their perception-surprisal.
Ultimately,
we are facing a Paretian statistical physics that challenge our Gaussian way of thinking.

The present work explores,
by hand,
how we can extend our Paretian statistical physics model for self-organized urban street network
to \textit{`nearly'} self-organized urban street networks.
Basically,
we want to proceed by applying arbitrary small perturbations to our model,
and see what we get.
In the remaining,
the paper is organized as follows.
The second section articulates the pace
from raw urban street networks
to idealized self-organized urban street networks
within the framework of information physics.
Next,
the third section shifts to Jaynes’s maximum entropy principle.
There,
along treatments and discussions,
we set
the idealized self-organized Shannon-Lagrangian for urban street networks
before we perturb it with an external surprisal drift.
Eventually,
after highlighting the two major practical properties of our theoretical work,
we point to future observational works
around the universality of self-organized urban street networks
as such and as reference.

\begin{figure}[bth!]
	\includegraphics[width=0.95\linewidth]{osusn_jsi-figure-01}
	\caption{\label{OSUSN/fig/USN/NotionalExample/RawMaterial}%
		Notional urban street network%
		\endnote{%
			Notional example inspired by the \textit{`notional road network'}
			in the paper by \citet*{BJiangSZhaoJYin2008}.%
			}\label{OSUSN/edn/NotionalExample}
		in black-and-white and colourized versions
		used all along the paper.
		This notional example
		is meant to
		pattern a
		portion of a real-world city map.
		The black-and-white version ($\mathrm{g}$) connotes a geometrical viewpoint
		that leads to a Poissonian physics.
		Whereas
		the colourized version ($\mathrm{t}$) evinces a topological perception
		that is subject to scale-free behaviours.
		}
\end{figure}

\section*{\addcontentsline{toc}{section}{From Apparent Dullness to Living Coherence}From Apparent Dullness to Living Coherence}

\subsection*{\addcontentsline{toc}{subsection}{Structure to Quantify}Structure to Quantify}

\subsubsection*{\addcontentsline{toc}{subsubsection}{From Street-Junction Networks to Road-Road Networks}\label{sec/subsub/AD2LC/S2Q}From Street-Junction Networks to Road-Road Networks}

\begin{figure}[bth!]
	\includegraphics[width=0.95\linewidth]{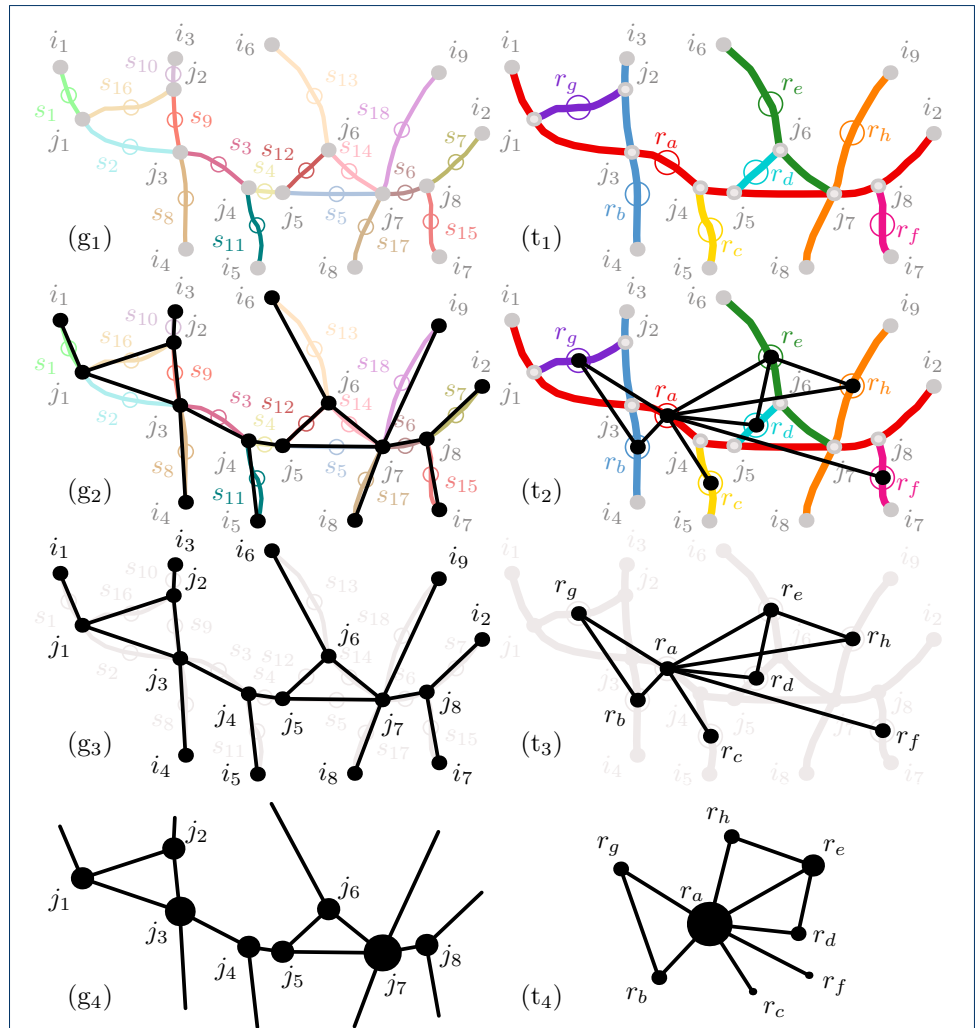}
	\caption{\label{OSUSN/fig/USN/NotionalExample/GeoTopo}%
		Geometrical versus topological approaches for urban street networks:
		a four-step visual construction of their respective abstract networks.
		Each construction is performed on the notional sample exhibited
		in Figure~\ref{OSUSN/fig/USN/NotionalExample/RawMaterial}.
		The left four-step sequence ($\mathrm{g}_{1}$)--($\mathrm{g}_{4}$)
		and its right counterpart ($\mathrm{t}_{1}$)--($\mathrm{t}_{4}$)
		sketch for this sample
		the geometrical and topological abstract network constructions,
		respectively.
		At Step~$1$, street-segments and roads are identified:
			the street-segments are labelled with indexed $s$ and coloured in distinct pallid colours;
			the roads are labelled with indexed $r$ and coloured in distinct vivid colours.
			Meanwhile,
			the junctions and the impasses are coloured in grey and labelled with indexed $j$ and $i$, respectively.
		In Subfigure~$\mathrm{g}_{2}$,
			the extended junctions $j_{\ast}$ and $i_{\ast}$ and the street-segments $s_{\ast}$
			spontaneously become nodes and edges, respectively.
		In Subfigure~$\mathrm{t}_{2}$,
			each road $r_{\ast}$ is reduced to a node
			and each road-node pair $\{r_{\ast},r_{\star}\}$ is linked
			whenever $r_{\ast}$ and $r_{\star}$ share
			at least
			a common junction.
		At Step~$3$, the raw material is being dissolved to highlight the emerging abstract networks.
		Finally, at Step~$4$,
			the resulting abstract networks are rearranged to stress their relevant traits:
			the size of each node is proportional to its valence;
			the impasses $i_{\ast}$ are neglected because they are rather free-ends than nodes;
			the road-node $r_{h}$ was flipped to avoid a confusing edge crossing;
			and so forth.
		}
\end{figure}

Everyone has seen black-and-white city maps drew with lines of the same width
as shown in Figure~\ref{OSUSN/fig/USN/NotionalExample/RawMaterial}$\mathrm{g}$.
Each line intersection represents a street-junction (or junction),
each portion of line between two adjacent junctions may be identified
as a street-segment.
Basically,
an urban street network is composed of junctions bonded by street-segments.
That is,
junctions and street-segments constitute,
respectively,
the immediate nodes and links of a family of real-world networks
known as urban street networks
---
see Figure~\ref{OSUSN/fig/USN/NotionalExample/GeoTopo}.
As such,
these real-world networks are literally street-junction networks.
Construction rules readily impose that each junction ties together at least three street-segments.
On the other hand,
everyday observations tell us that,
anywhere in any city,
any junction joins mostly four street-segments,
occasionally five or six, rarely seven, and very exceptionally more.
Real data analysis shows that
the valence distribution for street-junction networks
essentially follows a Poisson law sharply centred in four \citep{BJiangSZhaoJYin2008,BJiangCLiu2009}.
In this sense,
the complexity of street-junction networks tends to be as trivial as a regular square lattice.
This first attempt to describe urban street environments
---
better known as the \emph{geometrical approach}
---
may appear to be too naive
\citep{BJiangTAUSN2004,BJiangSZhaoJYin2008,BJiangCLiu2009,BJiangTSUSNPDC2014,PortaTNAUSPA2006,MRosvall2005,APMsucci2009}.

As an alternative,
we may consider instead colourized city maps with lines of arbitrary colours
as shown in Figure~\ref{OSUSN/fig/USN/NotionalExample/RawMaterial}$\mathrm{t}$.
We have in mind street maps.
Basically,
a street map of a city has the particularity to exhibit
how the city-dwellers perceive the urban street network of their own city.
Explicitly,
it shows how they have gathered
along the time
the street-segments of their own city to form streets.
Implicitly,
it reveals that we
human
townmen
rather reason in terms of streets than of street-segments.
But over all,
deeply,
it betrays a topological mindset
that looks on street maps
essentially
for topological information.
Indeed,
to move from one place to another,
we seek for directional information
with the following three characteristic traits:
\newcounter{counterTMSTEnum}\setcounter{counterTMSTEnum}{1}%
(\roman{counterTMSTEnum})\stepcounter{counterTMSTEnum}~%
each pair of successive streets must critically share a common junction
---
whichever it is;
(\roman{counterTMSTEnum})\stepcounter{counterTMSTEnum}~%
each junction in itself plays a secondary role;
(\roman{counterTMSTEnum})\stepcounter{counterTMSTEnum}~%
neither position nor distance is important.
The \emph{topological approach} forces these three characteristic traits
by reducing road maps to
(topological)
road-road networks.
Here a {natural road}
(or {road}, for short)
is an accepted substitute for street
(more precisely, for ``named'' street).
A road-road network reduces roads to nodes and bonds each pair that shares a common junction
---
see Figure~\ref{OSUSN/fig/USN/NotionalExample/GeoTopo}.
Real data analysis shows that
the valence distribution for the road-road network
of a self-organized urban street network
typically follows an inverse-power scaling law,
namely,
a scale-free power law
\citep{BJiangTAUSN2004,CrucittiCMSNUS2006,PortaTNAUSPA2006,PortaTNAUSDA2006,BJiangATPUSN2007,BJiangSZhaoJYin2008,BJiangTSUSNPDC2014}.
This is scale-freeness.
We have a slight grasp of scale-freeness
for an urban street network
whenever we apprehend that
only a few streets cross a large number of them,
several streets cross an intermediate number of them,
and very many streets cross a small number of them.
As a matter of fact,
by contrast to street-junction networks,
road-road networks are subject to complex network behaviours.

Thusly,
the topological approach appears far more pertinent
than the geometrical one
for at least two reasons.
Firstly,
the topological description unveils that
urban street networks underlie complex behaviours generally observed in real complex networks
\citep{BJiangSZhaoJYin2008,BJiangCLiu2009,BJiangTSUSNPDC2014,PortaTNAUSPA2006,MRosvall2005,APMsucci2009,APMsucci2016};
the complexity induced by the geometrical description is trivial \citep{BJiangSZhaoJYin2008,BJiangCLiu2009}.
Secondly,
the topological approach permits to isolate a category of real urban street networks that
shows evidence of a \textit{`pure'} scaling behaviour;
the geometrical approach renders all urban street networks equally \textit{`boring'} \citep{BJiangSZhaoJYin2008}.
This idealized category of urban street networks may serve as a reference from which
any general urban street networks deviates.

\subsubsection*{\addcontentsline{toc}{subsubsection}{Road-Road Networks Mask Road-Junction Partial Orders}Road-Road Networks Mask Road-Junction Partial Orders}

Even though the topological approach leads to precious observations,
it remains mostly a descriptive tool.
The topological approach does not provide any explanation,
it is not concerned about the underlying principles
for how urban street networks are emerging.
A \emph{structural approach} that does not bypass street-junctions
(or road-junctions)
allows us to establish a statistical physics foundation
for the \textit{`pure'} scaling behaviour
as effectively observed among self-organized urban street networks \citep{SESOPLUSN}.
It is fair to add that the structural approach may lead to alternative foundations,
but also that it does not fail to catch the \textit{`true structure'}
of urban street networks
by forcing the three above topological characteristic traits
a bit too early.

Here
urban street networks are envisioned as a whole
where road literally tie together through junctions.
To begin with,
we represent the ties by an incidence relation
that gathers for each road all junctions through which it passes \citep{BJiangSZhaoJYin2008}
as exemplified in Table~\ref{OSUSN/tab/USN/NaturalRoads/IncidenceRelation}.
Then,
we interpret this road-junction incidence relation
as an object/attribute relation
for which any road acts as an objects and any junction as an attribute
\citep{RHAtkin1974,BADaveyHAPriestleyILO,YSHoTPP1982D}.
Eventually,
by invoking the Formal Concept Analysis (\textsc{FCA}) paradigm,
this change of perspective allows us to establish bijectively a partial-order relation
\citep{BADaveyHAPriestleyILO,YSHoTPP1982D}.
In other words,
every urban street network
is subject to and bijectively representable by a partial-order.

\begin{table}[h!]
\colorlet{tblOPSOUSNGreyXCol}{lightgray}
\providecommand{\tblOPSOUSNCross}{$\CIRCLE$}%
\let\tblOPSOUSNGreyX\tblOPSOUSNCross
\providecommand{\tblOPSOUSNAntiX}{{\tiny{$\cdot$}}}%
\settowidth{\tabcolsep}{$\:$}
	\caption{\label{OSUSN/tab/USN/NaturalRoads/IncidenceRelation}%
		Road-junction incidence dot-chart
		associated to the colourized notional urban street network
		introduced in Figure~\ref{OSUSN/fig/USN/NotionalExample/RawMaterial}
		with the labelling chosen in Figure~\ref{OSUSN/fig/USN/NotionalExample/GeoTopo}.
		Here the incidence relation is represented as a Boolean array that stores its Boolean values:
		a big dot \tblOPSOUSNCross\ stands for \texttt{true}, a tiny dot \tblOPSOUSNAntiX\ for \texttt{false};
		each row represents a road $r_{\ast}$, each column a junction $j_{\ast}$;
		$\fcaIR$ denotes the incidence relation.
		Incidence relations are concretization of object-attribute relations.
		Here the objects are the roads $r_{\ast}$ while the attributes are the junctions $j_{\ast}$.
		}
	\begin{tabular}{l|ccccccccccccccccc}%
		$\fcaIR$ &
			$j_{1}$ & $j_{2}$ & $j_{3}$ & $j_{4}$ & $j_{5}$ & $j_{6}$ & $j_{7}$ & $j_{8}$ &%
			$i_{1}$ & $i_{2}$ & $i_{3}$ & $i_{4}$ & $i_{5}$ & $i_{6}$ & $i_{7}$ & $i_{8}$ & $i_{9}$ \\
		\hline
		$r_{a}$ & \tblOPSOUSNCross & \tblOPSOUSNAntiX & \tblOPSOUSNCross & \tblOPSOUSNCross &%
			\tblOPSOUSNCross & \tblOPSOUSNAntiX & \tblOPSOUSNCross & \tblOPSOUSNCross &%
			\tblOPSOUSNGreyX & \tblOPSOUSNGreyX & \tblOPSOUSNAntiX & \tblOPSOUSNAntiX &%
			\tblOPSOUSNAntiX & \tblOPSOUSNAntiX & \tblOPSOUSNAntiX & \tblOPSOUSNAntiX & \tblOPSOUSNAntiX \\
		$r_{b}$ & \tblOPSOUSNAntiX & \tblOPSOUSNCross & \tblOPSOUSNCross & \tblOPSOUSNAntiX &%
			\tblOPSOUSNAntiX & \tblOPSOUSNAntiX & \tblOPSOUSNAntiX & \tblOPSOUSNAntiX &%
			\tblOPSOUSNAntiX & \tblOPSOUSNAntiX & \tblOPSOUSNGreyX & \tblOPSOUSNGreyX &%
			\tblOPSOUSNAntiX & \tblOPSOUSNAntiX & \tblOPSOUSNAntiX & \tblOPSOUSNAntiX & \tblOPSOUSNAntiX \\
		$r_{c}$ & \tblOPSOUSNAntiX & \tblOPSOUSNAntiX & \tblOPSOUSNAntiX & \tblOPSOUSNCross &%
			\tblOPSOUSNAntiX & \tblOPSOUSNAntiX & \tblOPSOUSNAntiX & \tblOPSOUSNAntiX &%
			\tblOPSOUSNAntiX & \tblOPSOUSNAntiX & \tblOPSOUSNAntiX & \tblOPSOUSNAntiX &%
			\tblOPSOUSNGreyX & \tblOPSOUSNAntiX & \tblOPSOUSNAntiX & \tblOPSOUSNAntiX & \tblOPSOUSNAntiX \\
		$r_{d}$ & \tblOPSOUSNAntiX & \tblOPSOUSNAntiX & \tblOPSOUSNAntiX & \tblOPSOUSNAntiX &%
			\tblOPSOUSNCross & \tblOPSOUSNCross & \tblOPSOUSNAntiX & \tblOPSOUSNAntiX &%
			\tblOPSOUSNAntiX & \tblOPSOUSNAntiX & \tblOPSOUSNAntiX & \tblOPSOUSNAntiX &%
			\tblOPSOUSNAntiX & \tblOPSOUSNAntiX & \tblOPSOUSNAntiX & \tblOPSOUSNAntiX & \tblOPSOUSNAntiX \\
		$r_{e}$ & \tblOPSOUSNAntiX & \tblOPSOUSNAntiX & \tblOPSOUSNAntiX & \tblOPSOUSNAntiX &%
			\tblOPSOUSNAntiX & \tblOPSOUSNCross & \tblOPSOUSNCross & \tblOPSOUSNAntiX &%
			\tblOPSOUSNAntiX & \tblOPSOUSNAntiX & \tblOPSOUSNAntiX & \tblOPSOUSNAntiX &%
			\tblOPSOUSNAntiX & \tblOPSOUSNGreyX & \tblOPSOUSNAntiX & \tblOPSOUSNAntiX & \tblOPSOUSNAntiX \\
		$r_{f}$ & \tblOPSOUSNAntiX & \tblOPSOUSNAntiX & \tblOPSOUSNAntiX & \tblOPSOUSNAntiX &%
			\tblOPSOUSNAntiX & \tblOPSOUSNAntiX & \tblOPSOUSNAntiX & \tblOPSOUSNCross &%
			\tblOPSOUSNAntiX & \tblOPSOUSNAntiX & \tblOPSOUSNAntiX & \tblOPSOUSNAntiX &%
			\tblOPSOUSNAntiX & \tblOPSOUSNAntiX & \tblOPSOUSNGreyX & \tblOPSOUSNAntiX & \tblOPSOUSNAntiX \\
		$r_{g}$ & \tblOPSOUSNCross & \tblOPSOUSNCross & \tblOPSOUSNAntiX & \tblOPSOUSNAntiX &%
			\tblOPSOUSNAntiX & \tblOPSOUSNAntiX & \tblOPSOUSNAntiX & \tblOPSOUSNAntiX &%
			\tblOPSOUSNAntiX & \tblOPSOUSNAntiX & \tblOPSOUSNAntiX & \tblOPSOUSNAntiX &%
			\tblOPSOUSNAntiX & \tblOPSOUSNAntiX & \tblOPSOUSNAntiX & \tblOPSOUSNAntiX & \tblOPSOUSNAntiX \\
		$r_{h}$ & \tblOPSOUSNAntiX & \tblOPSOUSNAntiX & \tblOPSOUSNAntiX & \tblOPSOUSNAntiX &%
			\tblOPSOUSNAntiX & \tblOPSOUSNAntiX & \tblOPSOUSNCross & \tblOPSOUSNAntiX &%
			\tblOPSOUSNAntiX & \tblOPSOUSNAntiX & \tblOPSOUSNAntiX & \tblOPSOUSNAntiX &%
			\tblOPSOUSNAntiX & \tblOPSOUSNAntiX & \tblOPSOUSNAntiX & \tblOPSOUSNGreyX & \tblOPSOUSNGreyX \\
	\end{tabular}
\end{table}

\begin{figure}[bth!]
	\includegraphics[width=0.975\linewidth]{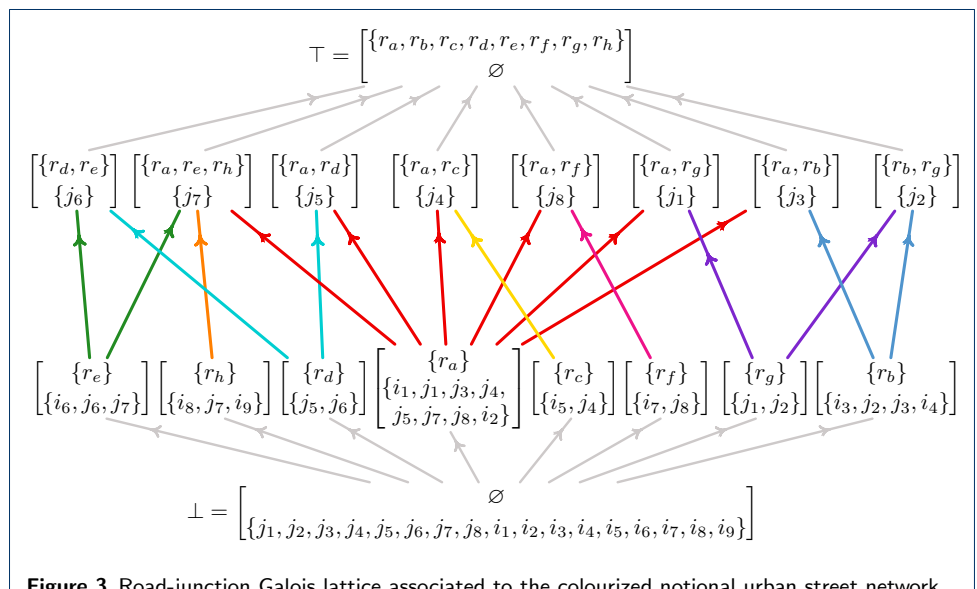}
	\caption{\label{OSUSN/fig/USN/NotionalExample/GL/HasseDiag}%
		Road-junction Galois lattice
		associated to the colourized notional urban street network
		introduced in Figure~\ref{OSUSN/fig/USN/NotionalExample/RawMaterial}
		with the labelling chosen in Figure~\ref{OSUSN/fig/USN/NotionalExample/GeoTopo}.
		This Galois lattice is obtained by applying the Formal Concept Analysis (\textsc{FCA}) paradigm
		to the incidence relation $\fcaIR$ whose chart representation is given
		in Table~\ref{OSUSN/tab/USN/NaturalRoads/IncidenceRelation}.
		This construction is one-to-one.
		A Galois lattice is an algebraic structure that underlies a partial order relation $\glprec$ and two algebraic operators,
		a join operator $\vee$ and a meet operator $\wedge$.
		The partial order relation can be interpreted as an extended logical imply relation $\glimplies$.
		The arrows in the diagram inherit this interpretation.
		For \textsc{FCA} lattices,
		each element is a pair of sets $[R,J]$ where $R$ is a set of objects and $J$ a set of attributes.
		Here the roads $r_{\ast}$ are the objects whose attributes are junctions $j_{\ast}$ and impasses $i_{\ast}$
		(see Table~\ref{OSUSN/tab/USN/NaturalRoads/IncidenceRelation}).
		Because the roads $r_{\ast}$ do not cross to each others more than once,
		the Galois lattice takes an intuitive two-layer form.
		Indeed,
		the join-irreducible elements $[\{r_{\ast}\},J]$ and the meet-irreducible elements $[R,\{j_{\ast}\}]$
		readily identify themselves with their road $r_{\ast}$ and their junctions $j_{\ast}$, respectively.
		So, the roads $r_{\ast}$ and the junctions $j_{\ast}$ immediately form, respectively,
		the lower and upper nontrivial layers of the Galois lattice.
		This also gives meaningful and intuitive interpretations to
		the partial order relation $\glprec$ and to the operators $\vee$ and $\wedge$:
		${r_{a}}\glprec{j_{7}}$
		(or ${r_{a}}\glimplies{j_{7}}$)
		reads ``road $r_{a}$ passes through junction $j_{7}$'' or ``junction $j_{7}$ is along road $r_{a}$'';
		${r_{a}}\vee{r_{b}}={j_{3}}$ reads ``roads $r_{a}$ and $r_{b}$ join at junction $j_{3}$'';
		${j_{3}}\wedge{j_{7}}={r_{a}}$ reads ``junctions $j_{3}$ and $j_{7}$ meet road $r_{a}$''.
		Each colourized arrow in the diagram bears the colour of its road.
		The top element $\top$ is the urban street network as a whole,
		while the bottom element $\bot$ is its absurd counterpart,
		emptiness or the absence of urban street network.
		}
\end{figure}

More interestingly,
any partial-order can be represented by an abstract ordered structure
known as Galois lattice \citep{BADaveyHAPriestleyILO,YSHoTPP1982D}.
In general,
a Galois lattice organizes itself in layers
with respect to its partial-order,
so that it can give rise to sympathetic graphical representations
called Hasse diagrams \citep{BADaveyHAPriestleyILO}.
The Galois lattice corresponding to the incidence relation in Table~\ref{OSUSN/tab/USN/NaturalRoads/IncidenceRelation}
is represented by a Hasse diagram in Figure~\ref{OSUSN/fig/USN/NotionalExample/GL/HasseDiag}.
If we assume that two roads cross to each other only once,
it appears then that
urban street networks reduce to intuitive two-layer Galois lattices:
the roads and the junctions make up the lower nontrivial layer and the upper nontrivial layer, respectively;
the \textit{`imply'} ordering relation
(or join operator)
is ``passing through'' (or ``crossing at'').
Figure~\ref{OSUSN/fig/USN/NotionalExample/GL/HasseDiag} exhibits clearly this property.
Roads that cross to each other more than once form loops.
Since such loops are rare while mostly not spontaneous,
for the sake of simplicity and unless otherwise specify,
the remaining will consider set of roads free of such loops.

Distributivity is an important property of Galois lattices \citep{BADaveyHAPriestleyILO}.
In particular,
for any finite Galois lattice,
distributivity allows us to claim that the elements of the first nontrivial layer
are the join-irreducible elements of the Galois lattice;
that is,
each upper elements can be expressed as
a join chain composed with elements of the first nontrivial layer
---
while none element of the first nontrivial layer can be decomposed.
In our context,
distributivity corresponds to the intuition that
any junction is a crossing of only two roads.
Therefore,
any junction that joins more than two roads
render the underlying Galois lattice nondistributive.
However,
as we have seen above,
junctions mostly join two roads.
Second,
any junction that joins at least three roads can be replaced by a roundabout
so that it remains only junctions that joins at most two roads.
For theses two reasons,
we may qualify as \emph{canonical}
any urban street network whose junctions effectively join only two roads.

Furthermore,
it is noticeable that
the underlying Galois lattice proves
not only to reduce bijectively but also to reflect pertinently
the involving topological complexity.
Indeed,
each underlying Galois lattice assigns
a clear primary role to roads
and a clear secondary role to junctions
so that the three topological characteristic traits are valorized as they should be:
roads imply junctions;
roads are join-irreducible (or just irreducible, for short),
while junctions are join-reducible (reducible) to roads;
junctions are meet-irreducible,
while roads are meet-reducible to junctions.
[%
As an aside,
for roads that form loops with each others,
the \textsc{FCA} paradigm simply creates abstractions of roads and junctions:
roads and junctions may then be defined,
respectively,
as the join-irreducible and meet-irreducible elements
of the involved road-junction Galois lattice.%
]
By now,
most of us should recognize road-road networks as
either zeroth order approximations or projections
of road-junction Galois lattices
by employing either an analytic analogy or a geometric one,
respectively.

To summarize,
any urban street network bijectively reduces
its topological complexity
to an essentially distributive two-layer Galois lattice,
while its canonicalization renders the latter plainly distributive.

\subsubsection*{\addcontentsline{toc}{subsubsection}{Partial Orders Are Algebraic Structures}Partial Orders Are Algebraic Structures}

Actually,
Galois lattices are not only ordered structures but also algebraic structures
\citep{BADaveyHAPriestleyILO,YSHoTPP1982D}.
To put it another way,
the join operator
(or partial order)
not only permit us
to construct the entire Galois lattice from its join-irreducible elements
\citep{BADaveyHAPriestleyILO,YSHoTPP1982D}
but also
to consistently assign numbers to its elements
so that the algebra of these numbers
reflects the algebra
of the Galois lattice
while their order respects its partial-order
\citep{KHKnuth2011,KHKnuth2008,KHKnuth2009,KHKnuth2005,KHKnuth2014}.
The evaluation of partially ordered sets
(or Galois lattices)
is the main object of the theory of information physics
\citep{KHKnuth2011,KHKnuth2008,KHKnuth2009,KHKnuth2005,KHKnuth2014}.
Because
quantifying roads and their concomitant junctions enable us
to confront models motivated by principles against observed data,
the structural approach liberates us to restrain ourselves
to perform sophisticated but nevertheless blind data analysis.
Given this,
the structural approach may no more appear as a gratuitous {\ae}sthetic step
to the most skeptical readers.
In brief,
the structural approach is a game changer.

In fact,
three Galois lattices are getting involved \citep{KHKnuth2008,KHKnuth2009}.
Let us now,
in order to forge for ourselves a better comprehensive picture,
succinctly describe them
and their respective \emph{valuation functions}.
The task is relatively easy since we are already familiar
with the Galois lattice of our system,
with the valuation function of the first extra Galois lattice,
and almost with the valuation function of the last Galois lattice.
Of course,
our first and foremost Galois lattice is the system itself,
so that our
unique
unknown valuation function $\glVa$ is simply meant to describe
the physics of our system.
Specifically,
the unknown valuation function $\glVa$ assigns a positive real number
indiscriminately
to all roads and junctions
so that each assigned positive real number characterises
the physical state of the involved road or junction.
Notice that valuation functions must be positive for consistency reasons.
As generic
Galois lattice
components,
roads and junctions organizes themselves in downsets.
A downset is a set of elements which contains all the elements implying each of them
\citep{BADaveyHAPriestleyILO}.
If we mentally sketch our urban street network randomly by roads and junctions
under the unique rule that a junction can be dotted only when all of its joining roads are already lined,
then each downset represents a state of our mental picture
---
and vice versa.
The set of all downsets ordered according to set inclusion $\subseteq$ forms a distributive Galois lattice,
which is called the \emph{state space}.
The state space is an auxiliary Galois lattice which merely helps us to introduce the next relevant one.

The join-irreducibles of the state space are the downsets associated to every road or junction,
that is,
the singleton sets composed of one road and sets composed of one junction along all of its joining roads.
These join-irreducibles generate the state space with set union $\cup$ as join-operator.
Nevertheless,
in reality,
given city-dwellers may not know precisely which state
their mental picture of the urban street network represents.
Even so,
they may have some information that exclude some states,
but not others.
Therefore,
the mental pictures of city-dwellers are mostly sets of potential states than single states know with certainty.
A set of potential states is called a \emph{statement}.
The set of all possible statements is simply the powerset generated from the set of all states.
Once ordered according to set inclusion $\subseteq$,
the set of all statements becomes a distributive Galois lattice
whose the join-irreducibles are the states.
This Galois lattice is known as the \emph{hypothesis space}.
Within the hypothesis space,
statements follows a logical deduction order as
each statement literally implies
(or is included in)
a statement with certainty.
The valuation function associated to any hypothesis space is recognized
to be a probability distribution.
So,
we are already very familiar with the algebra satisfied by the valuation functions associated to hypothesis spaces.
Among valuation functions associated to Galois lattices,
this algebra can be shown to be the only one possible
by imposing natural algebraic consistency restrictions.

Let us digress briefly to bring our attention back to the system valuation function $\glVa$:
as immediate consequence,
for canonical urban street networks,
the evaluation $\glVa(j(r,s))$ of a junction $j(r,s)$ joining a pair of roads $(r,s)$
must be the sum of the evaluations $\glVa(r)$ and $\glVa(s)$ of the joined roads $r$ and $s$,
respectively;
we have
\begin{equation}\label{eq/USN/Evaluation/constraint/addition}
	\glVa(j(r,s)) = \glVa(r) + \glVa(s)
	.
\end{equation}
End of digression.

Because the hypothesis state is essentially a representation of the system,
it is reasonable to claim that
its valuation function $\Pr$ must be related to the valuation function $\glVa$ of our system,
that is, to the physics of our system.
Meanwhile,
Rota theorem \citep[Thm.~1, Cor.~2]{GCRotaOCEC1971} asserts that,
for a finite distributive Galois lattice,
the valuation function is perfectly determined by the arbitrary values taken by its join-irreducibles.
In other words,
the valuation function $\Pr$ does not depend on the very structure of the hypothesis space;
rather,
it depends on the arbitrary values assigned to the join-irreducibles of the hypothesis space,
which are the states.
Accordingly,
the probability assigned to each state has to be
an arbitrary function of its evaluation by the valuation function $\glVa$;
this is a composition.
This arbitrary function interprets itself as a \textit{weight function} $\glWg$.
We read
\begin{equation}\label{eq/USN/HypothesisSpace/Evaluation/composition/Pr}
	\Pr = \glWg \circ \glVa
	.
\end{equation}
The weight function $\glWg$ constitutes our second unknown function.
The construction of the hypothesis space from the state space
corresponds technically to an exponentiation \citep{BADaveyHAPriestleyILO}.

The exponentiation of the hypothesis space brings up an \emph{inquiry space}.
The inquiry space is a distributive Galois lattice whose elements are \emph{questions}.
Thus, by construction, any question is a set of statements that answer it.
The quantification of the inquiry space leads to a measure,
coined \emph{relevance}.
In fact,
the inquiry space is Carrollian in the sense that it contains both
vain
(and fanciful)
and real questions.
A respond to a real question is a true state of our system.
To wit, a real question permits to know the configuration of our system exactly and without ambiguity.
A vain question can only lead to partial or ambiguous knowledge of the configuration.
The join chain of all the join-irreducible questions is the smallest real question,
it is called the \emph{central issue}.
The questions above the central issue form a Galois sublattice that contains all and only real questions.
The join-irreducible elements of the real Galois sublattice appears to partition their answers.
This property is reflected in the choice of the relevance
by coercing the relevance of a partition question to depend
on the probability of the greatest statements of its partitions.
This choice imposes the relevance to satisfy
the four natural properties of entropies \citep{JAczelBForteCTNg1974}.
This means that relevance is a generalized measure of information
with Shannon entropy as basis \citep{JAczelBForteCTNg1974}.
This is one of the major results of information physics.
The relevance of the central issue identifies itself with the entropy.
Therefore,
for canonical urban street networks,
the functional entropy $\glEta[\glVa,\glWg]$
takes the form
\begin{equation}\label{eq/USN/StructureEntropy}
	\glEta[\glVa,\glWg] =
		\!\sum_{r}\left(\glHm\circ\glWg\right)\left(\glVa(r)\right)
		+%
		\!\!\!\sum_{j(r,s)}\!\left(\glHm\circ\glWg\right)\left(\glVa(r)\!+\!\glVa(s)\right)
\end{equation}
where the first summation runs over the roads~$r$
and the second one over the junctions $j(r,s)$ joining the pair of roads $(r,s)$,
while $\glHm\colon{x}\mapsto-x\ln{x}$ is
the Shannon entropy function.
We will keep to express information measures in \textsf{nat} units.
For further details on the theory of information physics,
we refer the reader to the work of \citet{KHKnuth2011,KHKnuth2008,KHKnuth2009,KHKnuth2005,KHKnuth2014}.
For now,
we have enough material to step forward.

\subsection*{\addcontentsline{toc}{subsection}{Quantify to Organize}Quantify to Organize}

\subsubsection*{\addcontentsline{toc}{subsubsection}{From Galoisean Hierarchy to Paretian Coherence}\label{sec/subsub/Q2O/GH2PC}From Galoisean Hierarchy to Paretian Coherence}

Network data analysis shows that city-dwellers have
a topological perception of their urban street networks.
On the other hand,
the topology of urban street networks hides
a simple road-junction partial order that
bijectively reduces to intuitive two-layer Galois lattices.
The Galoisean hierarchy is intuitive in the sense that its join-operator expresses
our intuition that two roads join to form a junction.
Nonetheless this intuitive hierarchy leads to
two layers whose cardinality might be perceived as incommensurable.
Typical big cities count far more than several roads and junctions.
The apparent simplicity of the underlying Galois lattices is
the result of an algorithmic thought.
Nonetheless the Galoisean hierarchy is three-fold.
While the ordering and algebraic perspectives are respectively structural and operational,
the whole is measurable.
The underlying algebraic structure leads unambiguously to a unique quantification modulo
two unknown functions that we are free to choose.
These two unknown functions are of different nature.
The valuation function $\glVa$ assigns to each road or junction of the urban street network
a numerical quantity that characterizes its physical state.
The weight function~$\glWg$,
or more precisely its composition with the valuation function~$\glVa$
as expressed in \eqref{eq/USN/HypothesisSpace/Evaluation/composition/Pr},
allows us to assign to each mental picture of the urban street network
a numerical quantity that characterizes its perception among the city-dwellers.
This assignment is simply the probability distribution $\Pr$ of our system.
Ultimately all these mental pictures are surrounded by all sort of questions
whose pertinence can be measured.
The relevance of the most pertinent question is better known as the entropy of the system.
The most plausible probability $\Pr$,
that is,
the quantification which tends to represent at best
the perception of the city-dwellers for their own urban street network,
must also be the most relevant one.
In other words,
the most plausible probability $\Pr$ must maximize
the functional entropy \eqref{eq/USN/StructureEntropy} of their urban street network.
This is nothing other than Jaynes’s maximum entropy principle
\citep{KHKnuth2008,ETJaynes1957I,ETJaynes1957II,HKKesava2009,JNKapurHKKesava1992,ETJaynes1978SYLI}.
Thusly,
our physical content shifts from a algorithmic order to a fluctuating organization.

Roads and junctions indiscriminately yield \emph{our initial ignorance} \citep{ETJaynes1978SYLI}.
The most we can tell is that roads and junctions are mesoscopic systems
with a finite number of possible configurations $\Omega$.
Besides,
we must assume \emph{our complete ignorance} about their respective inner worlds.
This means that,
to our eyes at least,
all their possible configurations are equally likely.
Thusly,
roads and junctions are Boltzmannian mesoscopic systems.
Therefore,
the probability distribution $\Pr$ reduces to a function
that depends only on the number of possible configurations $\Omega$.
Meanwhile,
the functional entropy \eqref{eq/USN/StructureEntropy} simplifies
to take the more sympathetic form
\begin{equation}\label{eq/USN/StructureEntropy/Pr}
	\glEta[\Pr] =
		-\sum_{\Omega}\Pr(\Omega)\ln\left(\Pr(\Omega)\right)
	.
\end{equation}

On the other hand,
here,
self-organized urban street networks are idealized as scale-free systems,
\textit{viz.},
as systems exhibiting no typical number of configurations but rather a typical scale $\lambda$.
Thus,
as suitable characterizing moments
to invoke Jaynes’s maximum entropy principle \citep{ETJaynes1957I,ETJaynes1957II,HKKesava2009},
we must discard any classical moment and may consider logarithmic moments instead.
It appears that imposing the first logarithmic moment
\begin{equation}\label{eq/USN/ConfigurationSpace/Pr/FirstLogaritmicMoment/rhs}
	\sum\Pr(\Omega)\ln\Omega
\end{equation}
as sole characterizing constraint gives rise to
the scale-free probability distribution
\begin{equation}\label{eq/USN/ConfigurationSpace/Pr/approx}
	\Pr(\Omega)\propto\Omega^{-\lambda}
	.
\end{equation}
A practical normalization of this probability distribution leads to
the discrete Pareto probability distribution \citep{AClausetCRShaliziMEJNewman2009}.
To sum up:
the passage from the underlying Galoisean hierarchy
to an underlying Paretian coherence occurs
by invoking
Jaynes’s maximum entropy principle
with the first logarithmic moment as sole characterizing moment
and with our complete ignorance as initial knowledge condition.

For every road or junction having $\Omega$ possible configurations,
the Boltzmann entropy $\ln\Omega$ measures nothing but our complete ignorance
on the configuration effectively taking place.
So,
our characterizing restriction simply claims that
an idealized self-organized urban street network evolves
by preserving our complete ignorance on average.
This characterizing scheme that induces a Paretian coherence has been interpreted as
some evolutionary based mechanism
to maintain some opaque internal order \citep{MMilakovic2001,YDover2004}.
Note furthermore that \textit{``complete ignorance''} has rather remained,
so far,
a technical term.
A more intuitive interpretation might be considered instead.
If the Boltzmann entropy $\ln\Omega$ is interpreted as the \emph{surprisal} that
city-dwellers associate to every road or junction having $\Omega$ possible configurations,
then $\sum\Pr(\Omega)\ln\Omega$ becomes the amount of surprisal on average that
they associate to their own urban street network.
Surprisal
(or \emph{surprise})
$\stSu=-\ln\circ\Pr$
was introduced by \citet{MTribusTT} as a measure to quantify our astonishment and our indecision
whenever we face any arbitrary event.
Once adapted to our context,
surprisal somehow betrays the perception of the city-dwellers for their own urban street network.
Therefore,
the above Paretian characterizing constraint simply asserts that
an idealized self-organized urban street network evolves
by preserving on average the perception that its city-dwellers share for it.
This assertion renders city-dwellers the unconscious
but nevertheless active actors of their own urban street networks,
not the passive subjects of an obscure technical machinery.
Along this line,
the scale parameter $\lambda$ of the underlying scale-free probability distribution \eqref{eq/USN/ConfigurationSpace/Pr/approx}
interprets itself as an \emph{evolution scale}.

\subsubsection*{\addcontentsline{toc}{subsubsection}{Untangling the Underlying Coherence}Untangling the Underlying Coherence}

The underlying coherence,
Paretian or not,
does not reveal to city-dwellers as-is.
Technically,
we must still untangle the corresponding weight function $\glWg$ and valuation function $\glVa$
with respect to the underlying algebraic structure,
namely,
with respect to
composition \eqref{eq/USN/HypothesisSpace/Evaluation/composition/Pr}
and addition rule \eqref{eq/USN/Evaluation/constraint/addition}.
Practically,
we need a mesoscopic model to count the number of configurations $\Omega$ associated to every road or junction.
For the reason that
roads and junctions are likely driven by social interactions,
the mesoscopic model must typify social interactions.
To fulfill this purpose,
it appears convenient to adopt and adapt
the network of intraconnected agents model introduced by \citet{YDover2004}
for the distribution of cities in countries.
Thereby,
each road or junction becomes a hive of agents that connect to each other.
As agents,
we may consider the inhabitants that somehow
participate to the live activity of roads:
drivers, cyclists, pedestrians,
suppliers,
institutional agents,
residents,
and so forth.
For each road $r$,
the number of agents
is assumed to be asymptotically proportional to the number of junctions $n_{r}$
that $r$ crosses
--- the ratio $A$ being constant and sufficiently large.
This expresses nothing but the extensive property of roads.
Here the very existence of every road relies
on the ability for each of its agent
to maintain a crucial number of intraconnections
which is crudely equal to a constant number $\plNVCR$ \citep{YDover2004,RIMDunbarSShultz2007},
called the \emph{number of vital connections} for roads.
The layout of theses intraconnections is implicitly associated to
the internal order within each road,
while the total number of possible layouts
for each road
is simplistically considered as
its number of configurations \citep{YDover2004}.
\begin{subequations}\label{eq/USN/AgentBasedModel/NumberOfStates}
Therefore,
for each road~$r$,
the number of configurations $\Omega_{r}$ yields
\begin{equation}\label{eq/USN/AgentBasedModel/NumberOfStates/NaturalRoads}
	\Omega_{r}
		= \plNStR\left(n_{r}\right)
		\simeq
			\binom{\tfrac{1}{2}A\,n_{r}\left(A\,n_{r}-1\right)}{\plNVCR}
		\simeq
			\frac{A^{2\plNVCR}}{2^{\plNVCR}{\plNVCR}!}\,n_{r}^{2\plNVCR}
	.
\end{equation}
As concerns each junction,
continuing along this spirit,
the involved agents are merely
the agents of the two joining natural roads combined together.
Nevertheless,
as there is no apparent reason for roads and junctions to experience the same type of internal equilibrium,
we will assume two distinct numbers of vital connections,
$\plNVCR$ and $\plNVCJ$ respectively.
Then the same crude maneuvers give
\begin{equation}\label{eq/USN/AgentBasedModel/NumberOfStates/Junctions}
	\Omega_{j(r,s)}
		= \plNStJ\left(n_{j}=n_{r}+n_{s}\right)
		\simeq
			\frac{A^{2\plNVCJ}}{2^{\plNVCJ}{\plNVCJ}!}\,n_{j}^{2\plNVCJ}
	.
\end{equation}
\end{subequations}
Therefore,
the valuation function $\glVa$ appears clearly to assign
to each road or junction
the number of its agents,
and
the weight function $\glWg$ to asymptotically count
the number of possible vital intraconnection layouts
---
modulo normalization.

\section*{\addcontentsline{toc}{section}{Self-Organized Urban Street Networks as Reference}Self-Organized Urban Street Networks as Reference}

\subsection*{\addcontentsline{toc}{subsection}{Ideal Self-Organized Urban Street Networks}Ideal Self-Organized Urban Street Networks}

\subsubsection*{\addcontentsline{toc}{subsubsection}{Coherence Based on Boltzmannian Mesoscopic Surprisals}\label{sec/subsub/SOUSN/ISOUSN/derivation}Coherence Based on Boltzmannian Mesoscopic Surprisals}

It is time now to explicitly invoke Jaynes’s maximum entropy principle
for the functional entropy \eqref{eq/USN/StructureEntropy/Pr}
with the first logarithmic moment \eqref{eq/USN/ConfigurationSpace/Pr/FirstLogaritmicMoment/rhs}
as single characterizing constraint.
Promptly,
the corresponding Shannon Lagrangian writes
\begin{multline}%
\label{eq/USN/ShannonLagrangian}
	\mathcal{L}\left(\left\{\Pr(\Omega)\right\};\nu,\lambda\right) =
		-\sum_{\Omega} \Pr(\Omega)\,\ln\left(\Pr(\Omega)\right)
		-\left(\nu-1\right) \left[
				\sum_{\Omega} \Pr(\Omega) - 1
			\right]
		\\
		\shoveright{%
		\qquad\qquad\quad%
		-\lambda \left[
				\sum_{\Omega} \Pr(\Omega)\,\ln{\Omega} - {\left\langle{\mathrm{S}}\right\rangle}
			\right]
		.
		}
\end{multline}
The constraint relative to the Lagrange multiplier $\lambda$ compels
to keep constant the first logarithmic moment
\eqref{eq/USN/ConfigurationSpace/Pr/FirstLogaritmicMoment/rhs}
of the probability distribution $\Pr$;
namely,
it imposes
the preservation on average of the amount of surprisal
that city-dwellers perceive for their roads and junctions.
Meanwhile,
the Lagrange multiplier $\nu$ ensures the normalization condition that
the probability distribution $\Pr$ must satisfy.
The constant ${\left\langle{\mathrm{S}}\right\rangle}$ stands for
the constant mean surprisal at which the system evolves
---
for now it plays a dummy role.
Extremizing expression \eqref{eq/USN/ShannonLagrangian} yields
\begin{equation}\label{eq/USN/ShannonLagrangian/equations}
	\frac{\partial\mathcal{L}\left(\left\{\Pr\left(\Omega\right)\right\};\nu,\lambda\right)}{\partial\Pr(\Omega)} =
		-\ln\left(\Pr(\Omega)\right)
		-\nu
		-\lambda\,\ln\Omega
		= 0
	,
\end{equation}
which immediately leads to the scale-free probability distribution
\begin{equation}\label{eq/USN/ShannonLagrangian/solution/calculus/intermediate}
	\Pr(\Omega) =
		\frac{\Omega^{-\lambda}}{{e}^{\nu}}
\end{equation}
as previously claimed.
Afterwards,
the normalization condition effortlessly gives us an expression
for the dependent exponential denominator $\exp(\nu)$,
which may be defined as the \emph{partition function} $Z(\lambda)$ of our system;
we have
\begin{equation}
	{e}^{\nu} = \sum_{\Omega} \Omega^{-\lambda} \equiv Z(\lambda)
	.
\end{equation}
Ultimately,
we write solution \eqref{eq/USN/ShannonLagrangian/solution/calculus/intermediate}
in the more familiar form
\begin{equation}\label{eq/USN/ShannonLagrangian/solution}
	\Pr(\Omega) =
		\frac{\Omega^{-\lambda}}{Z(\lambda)}
	.
\end{equation}

The found most plausible probability distribution \eqref{eq/USN/ShannonLagrangian/solution}
concerns the underlying coherence of our system.
As such,
this coherence can only be perceived indirectly by the city-dwellers of the urban street network.
The city-dwellers may rather perceive the coherence behind roads and junctions.
Their corresponding statistics are obtained as follows.
Substituting \eqref{eq/USN/AgentBasedModel/NumberOfStates/NaturalRoads}
into \eqref{eq/USN/ShannonLagrangian/solution},
we readily obtain for roads
\begin{subequations}\label{eq/USN/AgentBasedModel/DistributionFunction}
\begin{equation}\label{eq/USN/AgentBasedModel/DistributionFunction/NaturalRoads}
	\Pr\left(n_{r}\right) \propto n_{r}^{-2\lambda\plNVCR}
	,
\end{equation}
which is a scale-free probability distribution.
Injecting instead \eqref{eq/USN/AgentBasedModel/NumberOfStates/Junctions}
into \eqref{eq/USN/ShannonLagrangian/solution},
then gathering and counting with respect to
the precedent probability distribution \eqref{eq/USN/AgentBasedModel/DistributionFunction/NaturalRoads}
gives for junctions
\begin{equation}\label{eq/USN/AgentBasedModel/DistributionFunction/Junctions}
	\Pr\left(n_{j}\right) \propto
		\left(
			\sum_{j(r,s)} \frac{\left[n_{j}=n_{r}+n_{s}\right]}{\left({n_{r}}{n_{s}}\right)^{2\lambda\plNVCR}}%
		\right)
		\,{n_{j}}^{-2\lambda\plNVCJ}
	,
\end{equation}
\end{subequations}
which is a generalized power law probability distribution;
the summation in parentheses is simply the self-convolution of
the road probability distribution \eqref{eq/USN/AgentBasedModel/DistributionFunction/NaturalRoads}.
The bracket around the equality statement follows Iverson's convention \citep{CONCMATH,DEKnuth1992}:
the bracket has value one whenever the bracketed statement is true, zero otherwise.
The number of junction $n_{r}$ that a road crosses is essentially the number of roads
with which it shares a common junction,
namely,
its valence number in the corresponding road-road network.
So the probability distribution \eqref{eq/USN/AgentBasedModel/DistributionFunction/NaturalRoads} predicts
the valence distribution for roads that has been widely observed empirically among self-organized cities
\citep{BJiangTAUSN2004,CrucittiCMSNUS2006,PortaTNAUSPA2006,PortaTNAUSDA2006,BJiangATPUSN2007,BJiangSZhaoJYin2008,BJiangTSUSNPDC2014}.
A similar argument dually applies for junctions.
However,
to the best of our knowledge,
the valence distribution for junctions has brought no attention until now
---
except in our recent investigations.

For practical data analysis \citep{AClausetCRShaliziMEJNewman2009},
we need to assume that
the number of junctions per road ${n}_{r}$ spans from some minimal positive value $\underline{n}_{r}$.
Then the normalization of probability distributions \eqref{eq/USN/AgentBasedModel/DistributionFunction}
can be performed elegantly
by using natural generalization of known special functions.
First,
the probability for a road to cross ${n}_{r}$ junctions becomes
\begin{subequations}\label{OSUSN/eq/USN/PDF}
\begin{equation}\label{OSUSN/eq/USN/PDF/NaturalRoads}
	\Pr\left({n}_{r}\right) =
		\frac{{n}_{r}^{-2\lambda\upsilon_{r}}}{\sfhzeta\left(2\lambda\upsilon_{r};\underline{n}_{r}\right)}
	,
\end{equation}
where
\begin{math}
	\sfhzeta\left(\alpha;a\right) =
		\sum_{{n}=0}^{\infty} {(a+n)}^{-\alpha}
\end{math}
is the generalized
(or Hurwitz-)
zeta function \citep[\S~25.11]{HBMF}.
Second,
the probability
for a junction to see ${n}_{j}$ junctions through its joining roads
reads
\begin{equation}\label{OSUSN/eq/USN/PDF/Junctions}
	\Pr\left({n}_{j}\right) =
		\frac{%
			\sum_{{n}=\underline{n}_{r}}^{{n}_{j}-\underline{n}_{r}}
				\left[{n}\left({n}_{j}-{n}\right)\right]^{-2\lambda\upsilon_{r}}
				\,%
				{n}_{j}^{-2\lambda\upsilon_{j}}
			}{%
			\sfWitten\left(2\lambda\upsilon_{r},2\lambda\upsilon_{r},2\lambda\upsilon_{j};\underline{n}_{r}\right)%
			}
	,
\end{equation}
where
\begin{math}
	\sfWitten\left(\alpha,\beta,\gamma;\underline{n}\right) =
		\sum_{{m},{n}\geqslant\underline{n}}
			{m}^{-\alpha} {n}^{-\beta} \left(m+n\right)^{-\gamma}
\end{math}
is the two-dimensional
generalized
(or Hurwitz-)
Mordell-Tornheim-Witten zeta function \citep{JMBorweinKDilcher2018}.
\end{subequations}

As a conclusion,
let us remark that
statistics \eqref{OSUSN/eq/USN/PDF} for an ideal self-organized urban street network
does not separate
the macroscopic parameter $\lambda$
from the mesoscopic ones $\plNVCR$ and $\plNVCJ$
in the sense that,
at best,
we can only estimate the products $\lambda\plNVCR$ and $\lambda\plNVCJ$.
This separation of parameters is critical since it would allow us to distinguish quantitatively
the macroscopic phenomenon of evolution
from the mesoscopic phenomena of social interactions
that take place in urban street networks.
Notice that,
from a qualitative perspective,
two distinct behaviours are anticipated.
The numbers of vital connections $\plNVCR$ and $\plNVCJ$ certainly differ
from one cultural basin to another one \citep{YDover2004}.
Whereas the evolution scale $\lambda$ might transcend cultures \citep{SCALE2017}.
A classical way to separate parameters in Physics consists to introduce sufficiently small perturbations.
This is,
in its observational form,
the subject of the next subsection.

\subsubsection*{\addcontentsline{toc}{subsubsection}{Case Study of Central London}\label{sec/subsub/CaseStudy/CentralLondon}Case Study of Central London}

Figure~\ref{OSUSN/fig/USN/London} shows
the Relative Frequency Distributions (\textsc{RFD}) of the urban street network of Central London.
The probability distribution for roads
$\Pr\left({n}_{r}\right)$ \eqref{OSUSN/eq/USN/PDF/NaturalRoads} appears highly plausible,
as expected for any recognized self-organized city \citep{CAlexanderACINAT1965,BJiangTSUSNPDC2014}.
However,
for the time being,
the validation of the probability distribution for junctions $\Pr\left({n}_{j}\right)$ \eqref{OSUSN/eq/USN/PDF/Junctions}
appears more delicate.
This is due to the emergence of a numerical bottleneck as follows.
The state-of-the-art statistical method to either validate or reject a plausible hypothesis
for power law probability distributions
is based on Maximum Likelihood Estimations
(\textsc{MLE}) \citep{AClausetCRShaliziMEJNewman2009}.
Besides invoking a numerical minimizer \citep{WHPress2007},
this method requires sampling \citep{AClausetCRShaliziMEJNewman2009},
that is,
the input sample must be compared to a large set of randomly generated samples
---
the larger, the more precise.
In the present case,
this means that the numerical evaluation of the normalizing functions $\sfhzeta$ and $\sfWitten$
---
and of their respective logarithms and logarithmic derivatives
---
have to be efficient not only in terms of precision but also in terms of speed.
Efficient numerical methods to evaluate the Hurwitz-zeta function $\sfhzeta$
can be found in the classical numerical literature
\citep{HBMF,KOldhamJCMylandJSpanier2009}
---
while they can easily be adapted to our specific usage.
By contrast,
the two-dimensional
Mordell-Tornheim-Witten zeta function $\sfWitten$
belongs to the specialized numerical literature and
its numerical computation is still a subject of investigation
\citep{JMBorweinKDilcher2018}.
In practice,
even the implementation of the corresponding Hurwitz generalization
with the same two first exponents
$\alpha$ and $\beta$
is rather tedious while very slow,
especially when the third exponent $\gamma$ becomes negative
---
as $2\lambda\upsilon_{j}$ appeared to be.
To work around this numerical bottleneck,
we performed a crude data analysis based on a Nonlinear Least-Squares Fitting
(\textsc{NLSF}).
Interestingly,
our {ad hoc} crude data analysis reveals
a negative number of vital connection $\upsilon_{j}$,
which means that the associated generalized binomial combination number is smaller than one
modulo a signed factor that drops at normalization%
\endnote{%
	We have
	\begin{math}
		\binom{N}{-\nu}
			= \frac{\sin\pi\nu}{\pi\nu}{\binom{N+\nu}{\nu}}^{-1}
			= \frac{\sin\pi\nu}{\pi\nu}{\binom{N}{\nu}}^{-1} (1+\mathcal{O}(\frac{\nu^2}{N}))
	\end{math}%
	.
	}\label{OSUSN/edn/Combinatorics}%
.
We interpret this to mean that
the number of intraconnections for junctions
might be relatively much smaller than the one for roads
in self-organized cities.

\begin{figure}[bth!]
	\includegraphics[width=0.85\linewidth]{osusn_jsi-figure-04}
	\caption{\label{OSUSN/fig/USN/London}%
		Relative Frequency Distributions (\textsc{RFD}) for the urban street network of Central London:
		circles represent relative frequencies for the valences of the road-road topological network;
		crosses represent relative frequencies for the valences of the junction-junction topological network.
		The red fitted curve
		for the natural road statistics
		describes the Maximum Likelihood Estimate (\textsc{MLE})
		for the discrete Pareto probability distribution \eqref{OSUSN/eq/USN/PDF/NaturalRoads}
		estimated according to the state of the art \citep{AClausetCRShaliziMEJNewman2009,CSGillespie2015}
		(%
			$\underline{n}_{r} = 4$,
			$2\lambda\upsilon_{r} = 2.610(65)$,
			$n=250\,000$ samples,
			$p\text{-value} = 0.933(1)$%
			).
		The green fitted curve for the junction statistics
		shows the best Nonlinear Least-Squares Fitting (\textsc{NLSF})
		for the nonstandard discrete probability distribution \eqref{OSUSN/eq/USN/PDF/Junctions}
		with $\underline{n}_r$ and $2\lambda\upsilon_{r}$ fixed to
		their respective \textsc{MLE} value ($2\lambda\upsilon_{j} \approx -1.3$);
		since
		fast evaluation of the normalizing function $\sfWitten$ has yet to be found,
		no \textsc{MLE} approach can be used for now.
		Having for junctions
		a number of vital connections $\upsilon_{j}$ negative
		is interpreted as expressing
		a number of agent intraconnections
		for junctions relatively much smaller than the one for natural roads.
		The sharp downturn at a valence of $10$ likely means that the model fails to catch
		what occurs when valences are small.
		In any case,
		a proper \textsc{MLE} remains to be performed for confirming.%
		}
\end{figure}

\subsection*{\addcontentsline{toc}{subsection}{Drifted Self-Organized Urban Street Networks}Drifted Self-Organized Urban Street Networks}

\subsubsection*{\addcontentsline{toc}{subsubsection}{Coherence Based on Drifted Boltzmannian Mesoscopic Surprisals}Coherence Based on Drifted Boltzmannian Mesoscopic Surprisals}

Now let us regard the self-organized urban street networks studied in the previous section
as an ideal class of urban street networks,
namely,
as a reference from which \textit{`real'} urban street networks deviate.
The deviation is vanishing for self-organized urban street networks.
For arbitrary urban street networks,
the deviation might be of arbitrary magnitude.
Furthermore,
we presume that deviations are essentially caused by artificial means,
but not due to any change in the behaviours of the city-dwellers.
Artificial deviations are created by urban designers or decision makers
who remodel cities for arbitrary purposes but without respect to the laws that
might govern the spontaneous evolution of cities.
Meanwhile,
the topological mindset of city-dwellers and the social machinery
that governs roads and junctions remain unchanged.
Moreover,
\textit{a priori},
there are no apparent reason that the remodelling
affects one iota
the deep paradigm which constructs the perception of city-dwellers:
roads and junctions remain perceived as Boltzmannian mesoscopic systems.
Nevertheless,
the remodeled urban street networks might no more reflect their perception
---
not vice versa.
In other words,
the deviations drift the surprisal of city-dwellers for their own urban street network.
Assuming a surprisal drift $\stDr(\Omega)$ that
generates an extra amount of surprisal ${\Delta{\left\langle{\mathrm{S}}\right\rangle}}$ on average,
the unique characterizing constraint bracket
in Shannon Lagrangian \eqref{eq/USN/ShannonLagrangian} becomes
\begin{equation}\label{eq/USN/drift/ShannonLagrangian/constraint/bracket/adhoc}
	\left[
		\sum_{\Omega}
			\Pr(\Omega)
				\left(%
					\ln{\Omega}+\stDr(\Omega)%
					\vphantom{\widetilde\Delta}%
				\right)
			-
			\left(%
				{\left\langle{\mathrm{S}}\right\rangle}+{\Delta{\left\langle{\mathrm{S}}\right\rangle}}%
				\vphantom{\widetilde\Delta}%
			\right)
	\right]
	.
\end{equation}
Carefully expanding \eqref{eq/USN/drift/ShannonLagrangian/constraint/bracket/adhoc}
gives rise to two apparent characterizing restrictions:
the first logarithmic moment characterizing constraint discussed above
and a new characterizing constraint,
respectively
\begin{equation}\label{eq/USN/drift/ShannonLagrangian/constraint/bracket/andsplit}
	\left[
		\sum_{\Omega}
			\Pr(\Omega)\,\ln(\Omega) - {\left\langle{\mathrm{S}}\right\rangle}
	\right]
	\quad\text{and}\quad
	\left[
		\sum_{\Omega}
			\Pr(\Omega)\,\stDr(\Omega) - {\Delta{\left\langle{\mathrm{S}}\right\rangle}}
	\right]
	.
\end{equation}
By adding this new characterizing restriction to Shannon Lagrangian \eqref{eq/USN/ShannonLagrangian},
we arrive at the deviant version
\begin{multline}%
\label{eq/USN/drift/ShannonLagrangian}
	\mathcal{L}\left(\left\{\Pr(\Omega)\right\};\nu,\lambda,\varepsilon\right) =
		-\sum_{\Omega} \Pr(\Omega)\,\ln\left(\Pr(\Omega)\right)
		-\left(\nu-1\right) \left[
				\sum_{\Omega} \Pr(\Omega) - 1
			\right]
		\\
		\shoveright{%
		\qquad\quad%
		-\lambda \left[
				\sum_{\Omega} \Pr(\Omega)\,\ln{\Omega} - {\left\langle{\mathrm{S}}\right\rangle}
			\right]
		-\varepsilon \left[
				\sum_{\Omega}
					\Pr(\Omega)\,\stDr(\Omega) - {\Delta{\left\langle{\mathrm{S}}\right\rangle}}
			\right]
		.
		}
\end{multline}
The introduced Lagrange multiplier $\varepsilon$ tells us how
urban designers or decision makers
impose
a surprisal drift $\stDr(\Omega)$ to the surprisal perception of city-dwellers
for their own urban street network.
The constant ${\Delta{\left\langle{\mathrm{S}}\right\rangle}}$ corresponds to
the part of the apparent mean surprisal caused by the surprisal drift $\stDr(\Omega)$ itself
---
for now,
as the constant ${\left\langle{\mathrm{S}}\right\rangle}$,
it plays a dummy role.
Extremizing expression \eqref{eq/USN/drift/ShannonLagrangian} holds
\begin{equation}\label{eq/USN/drift/ShannonLagrangian/equations}
	\frac{\partial\mathcal{L}\left(\left\{\Pr\left(\Omega\right)\right\};\nu,\lambda,\varepsilon\right)}{\partial\Pr(\Omega)} =
		-\ln\left(\Pr(\Omega)\right)
		-\nu
		-\lambda\,\ln\Omega
		-\varepsilon\,\stDr(\Omega)
		= 0
	,
\end{equation}
from which we readily find the power law probability distribution
\begin{equation}\label{eq/USN/drift/ShannonLagrangian/solution/calculus/intermediate}
	\Pr(\Omega) =
		\frac{\Omega^{-\lambda}\:{e}^{-\stDr(\Omega)}}{{e}^{\nu}}
	.
\end{equation}
With the same easy manipulation as before,
the normalization condition allows us to define
the deviant partition function $Z(\Lambda;\lambda,\varepsilon)$ of our drifted system;
we get
\begin{equation}
	{e}^{\nu} = \sum_{\Omega} \Omega^{-\lambda}\:{e}^{-\varepsilon\stDr(\Omega)} \equiv Z(\Lambda;\lambda,\varepsilon)
	.
\end{equation}
So we end up by writing the most plausible probability distribution
associated to Shannon Lagrangian \eqref{eq/USN/drift/ShannonLagrangian} as
\begin{equation}\label{eq/USN/drift/ShannonLagrangian/solution}
	\Pr(\Omega) =
		\frac{\Omega^{-\lambda}\:{e}^{-\varepsilon\stDr(\Omega)}}{Z(\Lambda;\lambda,\varepsilon)}
	.
\end{equation}
For non-vanishing surprisal drift $\varepsilon\stDr(\Omega)$,
as expected,
this probability distribution is obviously not scale-free.
In fact,
when the polynomial part of the asymptotic expansion of $\Lambda(\Omega)$ does not reduce to a constant,
the surprisal drift $\varepsilon\stDr(\Omega)$ acts as a cut-off function.
In other words,
in contrast to ideal self-organized urban street networks,
a typical deviant urban street network possesses a typical number of configurations
for its roads and junctions.

Our next task is to establish the statistics for roads and junctions in deviant urban street networks.
Substitution of \eqref{eq/USN/AgentBasedModel/NumberOfStates/NaturalRoads}
into \eqref{eq/USN/ShannonLagrangian/solution} yields
\begin{subequations}\label{eq/USN/drift/AgentBasedModel/DistributionFunction}
\begin{equation}\label{eq/USN/drift/AgentBasedModel/DistributionFunction/NaturalRoads}
	\Pr\left(n_{r}\right) \propto
		n_{r}^{-2\lambda\plNVCR}\:\exp\left({-\varepsilon\plDrR(n_{r}^{2\plNVCR})}\right)
	,
\end{equation}
once the surprisal drift $\stDr$ is suitably rescaled to the surprisal drift for roads $\plDrR$.
Afterwards,
substitution of \eqref{eq/USN/AgentBasedModel/NumberOfStates/Junctions}
into \eqref{eq/USN/ShannonLagrangian/solution}
along Iversonian counting
with respect to \eqref{eq/USN/drift/AgentBasedModel/DistributionFunction/NaturalRoads}
gives
\begin{multline}%
\label{eq/USN/drift/AgentBasedModel/DistributionFunction/Junctions}
	\Pr\left(n_{j}\right) \propto
		\left(
			\sum_{j(r,s)} \frac{\left[n_{j}=n_{r}+n_{s}\right]}{\left({n_{r}}{n_{s}}\right)^{2\lambda\plNVCR}}%
			\exp\left({-\varepsilon\left[\plDrR(n_{r}^{2\plNVCR})+\plDrR(n_{s}^{2\plNVCR})\right]}\right)
		\right)
		\\
		\times%
		{n_{j}}^{-2\lambda\plNVCJ}\:\exp\left({-\varepsilon\plDrJ(n_{j}^{2\plNVCJ})}\right)
	,
\end{multline}
\end{subequations}
with the same notation convention previously used.

The main interest of the deviant statistics \eqref{eq/USN/drift/AgentBasedModel/DistributionFunction} lies in showing
how surprisal drift formally separates
the evolution scale exponent $\lambda$
from the numbers of vital connections for roads and junctions,
$\plNVCR$ and $\plNVCJ$ respectively.
As seen in the previous subsection,
this separation of parameters is important as it means that
the macroscopic phenomenon of evolution
and the mesoscopic phenomena of social interactions
can be qualitatively studied
among drifted self-organized urban street networks.
Fortunately enough,
such qualitative investigations
among slightly drifted self-organized urban street networks
appears almost as manageable as the ideal case investigation
among self-organized urban street networks
as follows.

\subsubsection*{\addcontentsline{toc}{subsubsection}{Exploratory Study of Slightly Drifted Urban Street Networks}\label{sec/subsub/ExploratoryStudy/SDUDN}Exploratory Study of Slightly Drifted Urban Street Networks}

Let us first specify what we mean when a self-organized urban street network is slightly drifted.
Here it is important to bear in mind that
the numbers of configurations \eqref{eq/USN/AgentBasedModel/NumberOfStates}
result from asymptotic countings.
So,
the surprisal drifts for roads and junctions
introduced in \eqref{eq/USN/drift/AgentBasedModel/DistributionFunction},
$\plDrR$ and $\plDrJ$ respectively,
reach the asymptotic behaviour of the underlying surprisal drift $\stDr$.
Let us now assume that the underlying surprisal drift $\stDr(\Omega)$ admits
as asymptotic expansion a generic finite Laurent polynomial of the form
\begin{math}
	a_{-p}\Omega^{-p}+a_{-p+1}\Omega^{-p+1}+\cdots%
	+a_{0}+\cdots%
	+a_{q-1}\Omega^{q-1}+a_{q}\Omega^{q}%
\end{math}%
. 
The non-polynomial part is absorbed by the exponential function whose the surprisal drifts feed,
hence irrelevant.
The zeroth order coefficient $a_{0}$ is eliminated
during the normalization by factorizing its inverse exponentiation,
hence meaningless.
The remaining polynomial part
\begin{math}
	a_{1}\Omega+\cdots%
	+a_{q-1}\Omega^{q-1}+a_{q}\Omega^{q}%
\end{math}
is imposed,
by the normalization condition,
to be positive for large $\Omega$ values.
More importantly,
the remaining polynomial behaves as an asymptotic cut-off polynomial
whose strength lies in its leading term $a_{q}\Omega^{q}$.
We may consider surprisal drifts asymptotic to quadratic or of higher degree polynomials
as inducing too drastic cut-offs,
namely,
as altering too drastically self-organized urban street networks.
That is,
for the time being,
we consider as slight any surprisal drift that is asymptotic to
a monomial of degree one $a_{1}\Omega$ whose coefficient $a_{1}$ is arbitrary small
---
and positive.
Thus,
we may assume,
without loss of generality,
that the slight surprisal drift $\stDr(\Omega)$ reduces to the canonical monomial $\Omega$
so that $\varepsilon\stDr(\Omega)=\varepsilon\,\Omega$.
Therefore
the parameter $\varepsilon$ simply expresses the strength of our slight surprisal drift.
Once properly rescaled,
the parameter $\varepsilon$ becomes the strengths $\plDrSR$ and $\plDrSJ$ associated to
the slight surprisal drifts for roads and junctions,
respectively;
we write
\begin{subequations}\label{eq/USN/drift/AgentBasedModel/SurprisalDrift/slight}
\begin{align}
	\varepsilon\plDrR(n_{r}^{2\plNVCR}) &= \plDrSR\:{n_{r}^{2\plNVCR}}%
		\label{eq/USN/drift/AgentBasedModel/SurprisalDrift/slight/NaturalRoads}%
	\\
	\varepsilon\plDrJ(n_{j}^{2\plNVCJ}) &= \plDrSJ\:{n_{j}^{2\plNVCJ}}
		\label{eq/USN/drift/AgentBasedModel/SurprisalDrift/slight/Junctions}%
	.
\end{align}
\end{subequations}

Now we may carry out the statistics for roads and junctions in slightly deviant urban street networks.
Substituting \eqref{eq/USN/drift/AgentBasedModel/SurprisalDrift/slight}
into \eqref{eq/USN/drift/AgentBasedModel/DistributionFunction},
then making the change of parameters
\begin{equation}\label{eq/USN/drift/AgentBasedModel/SurprisalDrift/slight/ChangeOfParameters}
	\plZedR = \exp(-\plDrSR)
	\qquad
	\plZedJ = \exp(-\plDrSJ)
\end{equation}
for conciseness,
we obtain
\begin{subequations}\label{eq/USN/drift/AgentBasedModel/DistributionFunction/slight}
\begin{align}
	\Pr\left(n_{r}\right) &\propto
		{n}_{r}^{-2\lambda\plNVCR}\:{\plZedR}^{{n}_{r}^{2\plNVCR}}%
			\label{eq/USN/drift/AgentBasedModel/DistributionFunction/slight/NaturalRoads}%
	\\
	\Pr\left(n_{j}\right) &\propto
		\left(
			\sum_{j(r,s)} \frac{\left[n_{j}=n_{r}+n_{s}\right]}{\left({n_{r}}{n_{s}}\right)^{2\lambda\plNVCR}}%
			\:%
			{\plZedR}^{n_{r}^{2\plNVCR}} {\plZedR}^{n_{s}^{2\plNVCR}}
		\right)
		\,{n_{j}}^{-2\lambda\plNVCJ}\:%
		{\plZedJ}^{n_{j}^{2\plNVCJ}}
			\label{eq/USN/drift/AgentBasedModel/DistributionFunction/slight/Junctions}%
	.
\end{align}
\end{subequations}

For practical normalization,
the change of parameters \eqref{eq/USN/drift/AgentBasedModel/SurprisalDrift/slight/ChangeOfParameters}
appears
precious
to easily identify the involved special functions.
First,
the probability \eqref{OSUSN/eq/USN/PDF/NaturalRoads}
for a road to cross ${n}_{r}$ junctions
in an idealized self-organized urban street network
takes
in a slightly deviant urban street network
the form
\begin{subequations}\label{OSUSN/eq/USN/drift/PDF/slight}
\begin{equation}\label{OSUSN/eq/USN/drift/PDF/slight/NaturalRoads}
	\Pr\left({n}_{r}\right) =
		\frac{{n}_{r}^{-2\lambda\plNVCR}\:{\plZedR}^{{n}_{r}^{2\plNVCR}}}%
			{\sfLerch\left(\plZedR,2\lambda\upsilon_{r},\underline{n}_{r};2\plNVCR\right)}
	,
\end{equation}
where
\begin{math}
	\sfLerch\left(z,\alpha,a;\beta\right) =
		\sum_{{n}=0}^{\infty} {(a+n)}^{-\alpha} {z}^{{(a+n)}^\beta}
\end{math}
is the generalization introduced by \citet{BRJohnson1974} of
the Lerch transcendent function
\begin{math}
	\sfLerch\left(z,\alpha,a\right) =
		\sum_{{n}=0}^{\infty} {(a+n)}^{-\alpha} {z}^{{n}}
\end{math}
\citep[\S~25.14]{HBMF}.
Second,
the concomitant probability \eqref{OSUSN/eq/USN/PDF/Junctions}
for a junction to see ${n}_{j}$ junctions through its joining roads
then transforms into
\begin{equation}\label{OSUSN/eq/USN/drift/PDF/slight/Junctions}
	\Pr\left({n}_{j}\right) =
		\frac{%
			\sum_{{n}=\underline{n}_{r}}^{{n}_{j}-\underline{n}_{r}}
				\left[{n}\left({n}_{j}-{n}\right)\right]^{-2\lambda\upsilon_{r}}
				{\plZedR}^{n^{2\plNVCR}} {\plZedR}^{\left({n}_{j}-{n}\right)^{2\plNVCR}}
				\;%
				{n}_{j}^{-2\lambda\upsilon_{j}} {\plZedJ}^{n_{j}^{2\plNVCJ}}
			}{%
			\sfWittenLerch\left(%
				[\plZedR,\plZedR,\plZedJ],%
				[2\lambda\upsilon_{r},2\lambda\upsilon_{r},2\lambda\upsilon_{j}];%
				\underline{n}_{r};%
				[2\upsilon_{r},2\upsilon_{r},2\upsilon_{j}]
				\right)%
			}
	,
\end{equation}
where
\begin{equation*}
	\sfWittenLerch\left(%
			[x,y,z],%
			[\alpha,\beta,\gamma];%
			\underline{n};%
			[\iota,\kappa,\mu]%
		\right)
		=
		\!\!\sum_{{m},{n}\geqslant\underline{n}}
			{m}^{-\alpha} {n}^{-\beta} \left(m+n\right)^{-\gamma}
			{x}^{{m}^{\iota}} {y}^{{n}^{\kappa}} {z}^{{(m+n)}^{\mu}}
\end{equation*}
is introduced for the sake of completeness.
\end{subequations}

The validation of the slightly deviant statistics \eqref{OSUSN/eq/USN/drift/PDF/slight}
is more challenging than of the ideal statistics \eqref{OSUSN/eq/USN/PDF}
from which it deviates
for at least two reasons.
Firstly,
it is rather an exploratory work since we have no catalogue
of slightly deviant urban street networks from which we can pick up relevant samples.
Secondly,
the involved normalizing functions $\sfLerch$ and $\sfWittenLerch$ are both computationally challenging.
Nonetheless,
the numerical bottleneck is
again
a bearer of both good news and bad news.

The bad news,
without surprise,
is that further investigations on
the deviant probability distribution
$\Pr\left({n}_{j}\right)$ \eqref{OSUSN/eq/USN/drift/PDF/slight/Junctions}
for junctions
must be postponed too.
This is simply because
its normalizing function $\sfWittenLerch$ combines
together the difficulties inherited from the normalizing functions $\sfWitten$ and $\sfLerch$ while,
for the least,
a fast numerical evaluation for the former has yet to be found.
The good news is that
a very efficient numerical evaluation already exists for the latter.
In fact,
this numerical evaluation was presented with the Lerch transcendent function
as illustration \citep{SVAksenov2003}.
Formally,
it consists to apply to the series
a condensation transformation followed by a Levin d-transformation \citep{WHPress2007,CBrezinskiMRedivoZaglia1991}.
Technically,
its adaptation to the generalized Johnson-Lerch transcendent function $\sfLerch$ is straightforward.
In practice,
a careful implementation written in \texttt{C} language
that uses the Levin transformation encoded in the \texttt{HURRY} procedure \citep[Algo.~602]{TFessler1983}
as implemented in the GNU Scientific Library \citep{GSL}
appears efficient in terms of both precision and speed.

\section*{\addcontentsline{toc}{section}{Conclusions and Future Works}Conclusions and Future Works}

The primary goal of our investigation is
to understand the statistics of urban street networks.
The objective of this research was two-fold.
First,
see how our recent results on self-organized urban street networks
can be broaden to \textit{`nearly'} self-organized urban street networks.
Second,
learn what we can expect from this extension of our initial domain of investigation.
The implicit idea behind this approach is that most urban street networks
can be envisaged as a perturbation of a self-organized urban street network.

To start,
we present the surprisal statistical physics model that
we showed to govern self-organized urban street networks.
Afterwards,
by hand,
we perturb the model by introducing a surprisal drift.
We argue that the surprisal drift essentially results
from artificial remodellings imposed by urban designers or decision makers.
We obtain the generic statistics for arbitrarily drifted self-organized urban street networks,
and most importantly,
a practical statistics for the slightly drifted ones among them.
All along we learn two important and practical properties.
First,
as expected whenever any perturbation occurs,
surprisal drift perturbations lead to a separation of parameters.
Here,
perturbations separate the macroscopic evolution scale parameter
from the mesoscopic social interaction parameters.
Second,
data analysis for validating the practical statistics
for slightly drifted self-organized urban street networks
remains manageable
---
modulo some numerical analysis efforts.

Future works must
first and foremost
validate the practical statistic for slightly drifted self-organized urban street networks
for a sufficiently large bunch of \textit{`real'} urban street networks.
Thereafter,
the macroscopic and mesoscopic parameters must be estimated
for a representative set of slightly drifted self-organized urban street networks
in view to perform an observational qualitative investigation
on the involved phenomena.
This investigation is helpful to determine whether or not
the macroscopic phenomena of evolution and the mesoscopic phenomena of social interactions
transcend cultural basins.
There exists evidences that the latters are cultural dependent.
We believe,
in contrast,
that the macroscopic phenomena of evolution is characterized
by an universal constant evolution scale that might reflect either
spacial spanning,
unconscious processes,
or both;
if so,
an observational estimation must be isolated
and ultimately some rational must be found.
This sequence of observational investigations aims to confirm
for self-organized urban street networks the status of reference
among urban street networks.
Once confirmed,
it makes sense to compare the strengths of the surprisal drifts among
a representative set of urban street networks.
Because these strengths reflect the rational thoughts of urban designers and decision makers,
we expect to observe a random
(not to say irrational)
set of data.
On the other hand,
since surprisal drifts might be perceived  as a source of stress by city-dwellers,
these strengths might have correlation with data that mark
their disengagement for their own urban street networks.
If such correlations are effectively observed,
then the strength of surprisal drifts might be interpreted as a indicator of their disengagement,
namely,
as a quality measure of their city.

\printendnotes*[OPSOUSN]

\begin{backmatter}

\section*{Abbreviations}

\textsc{FCA}: Formal Concept Analysis;
\textsc{MLE}: Maximum Likelihood Estimate;
\textsc{NLSF}: Nonlinear Least-Squares Fitting;
\textsc{RFD}: Relative Frequency Distribution.

\section*{Availability of data and materials}

The datasets generated and analysed during the current study are available
from the corresponding author on reasonable request.

\section*{Author's contributions}

{JGMB}
	conceived and designed the study,
	programmed the map treatment/analysis tools,
	collected and treated the map data,
	performed the statistical analysis,
	and
	wrote the manuscript.
{SEGJ}
	helped to shape the manuscript.
Both authors read and approved the final manuscript.

\section*{Competing interests}

The authors declare that they have no competing interests.

\bibliographystyle{spbasic}

\end{backmatter}

\end{document}